\documentclass[aip,amsmath,amssymb,reprint]{revtex4-1}

\usepackage{graphicx}% Include figure files
\usepackage{dcolumn}% Align table columns on decimal point
\usepackage{bm}% bold math

\usepackage[utf8]{inputenc}
\usepackage[T1]{fontenc}
\usepackage{mathptmx}
\usepackage{etoolbox}
\usepackage{soul}

\usepackage{lipsum}

\usepackage{xcolor}
\usepackage[utf8]{inputenc}

\newcommand{\be}{\begin{equation}}
\newcommand{\ee}{\end{equation}}
\newcommand{\bea}{\begin{eqnarray}}
\newcommand{\eea}{\end{eqnarray}}
\newcommand{\Id}[1] {\int \! \! {\rm d}^3 #1}
\newcommand{\vr} {{\bf r}}
\newcommand{\vR} {{\bf R}}
\newcommand{\nn} {\nonumber}

\makeatletter
\def\@email#1#2{%
 \endgroup
 \patchcmd{\titleblock@produce}
  {\frontmatter@RRAPformat}
  {\frontmatter@RRAPformat{\produce@RRAP{*#1\href{mailto:#2}{#2}}}\frontmatter@RRAPformat}
  {}{}
}%
\makeatother

\begin{document}

\title{What can lattice DFT teach us about real-space DFT?}

\author{Nahual Sobrino}
\affiliation{Nano-Bio Spectroscopy Group and European Theoretical Spectroscopy
  Facility (ETSF), Departamento de Pol\'imeros y Materiales Avanzados: F\'isica, Qu\'imica y
  Tecnolog\'ia, Universidad del Pa\'\i s Vasco UPV/EHU, Avenida Tolosa 72, E-20018
  San Sebasti\'an, Spain}
\author{David Jacob}
\affiliation{Nano-Bio Spectroscopy Group and European Theoretical Spectroscopy
  Facility (ETSF), Departamento de Pol\'imeros y Materiales Avanzados: F\'isica, Qu\'imica y
  Tecnolog\'ia, Universidad del Pa\'\i s Vasco UPV/EHU, Avenida Tolosa 72, E-20018
  San Sebasti\'an, Spain}
\affiliation{IKERBASQUE, Basque Foundation for Science, Plaza Euskadi 5, E-48009
  Bilbao, Spain}
\author{Stefan Kurth$^{*,}$}
\email[ Corresponding author: ]{stefan.kurth@ehu.es}
\affiliation{Nano-Bio Spectroscopy Group and European Theoretical Spectroscopy
  Facility (ETSF), Departamento de Pol\'imeros y Materiales Avanzados: F\'isica, Qu\'imica y
  Tecnolog\'ia, Universidad del Pa\'\i s Vasco UPV/EHU, Avenida Tolosa 72, E-20018
  San Sebasti\'an, Spain}
\affiliation{IKERBASQUE, Basque Foundation for Science, Plaza Euskadi 5, E-48009
  Bilbao, Spain}
\affiliation{Donostia International Physics Center (DIPC), Paseo Manuel de
  Lardizabal 4, E-20018 San Sebasti\'an, Spain}

\date{\today}

\begin{abstract}
  In this paper we establish a connection between density functional theory (DFT) for
  lattice models and common real-space DFT. We consider the lattice DFT description of a
  two-level model subject to generic interactions in Mermin's DFT formulation in the grand
  canonical ensemble at finite temperature. The case of only density-density and Hund's rule
  interaction studied in earlier work is shown to be equivalent to an exact-exchange description
  of DFT in the real-space picture. In addition, we also include the so-called pair-hopping
  interaction which can be treated analytically and, crucially, leads to non-integer occupations of
  the Kohn-Sham levels even in the limit of zero temperature. 
  Treating the hydrogen molecule in a minimal basis is shown to be equivalent to our
  two-level lattice DFT model. By means of the fractional occupations of the KS orbitals (which, in
  this case, are identical to the many-body ones) we reproduce the results of full configuration
  interaction, even in the dissociation limit and without breaking the spin symmetry.
  Beyond the minimal basis, we embed our HOMO-LUMO model into a standard DFT calculation and,
  again, obtain results in overall good agreement with exact ones without the need of breaking
  the spin symmetry.
\end{abstract}

\maketitle

\section{Introduction}
\label{intro}

Density Functional Theory (DFT) is an in principle exact framework for solving the quantum many-body problem
of interacting electrons.\cite{hohenberg1964inhomogeneous} The Kohn-Sham (KS) formulation of DFT allows to cast
the problem into the particularly simple form of a system of effectively non-interacting electrons moving in a
mean-field potential.\cite{Kohn1965self} Owing to its computational efficiency, conceptual simplicity and
in many cases high accuracy, KS-DFT has become one of the cornerstones of electronic structure
theory and is now the standard
tool for the description of electronic matter in computational condensed matter physics, material science and
chemistry.\cite{ParrYang:89,DreizlerGross:90,burke2012perspective}

One of the challenges for KS-DFT is the accurate description of so-called strongly
correlated systems, for which the physics is dominated by the effects of strong
electron-electron interactions. Important examples of these systems are transition-metal
oxides, rare-earth compounds and transition-metal complexes.
In these systems strong electronic correlations lead to very diverse phenomena such
as the Mott metal-insulator transition, high-temperature superconductivity 
and spin-crossover
behavior.\cite{imada1998metal,lee2006doping,Guetlich:BJOC:2013,Ahn:NMat:2021}
One of the problems standard approximations to DFT struggle with is the proper
description of multi-determinant, open-shell ground states within
the KS framework which by construction is
single-determinant.\cite{cohen2012challenges,SuXu:17,PerdewRuzsinszkySunNepalKaplan21}
To address this problem, various extensions to DFT have been 
proposed, such as the DFT+U method\cite{anisimov1991band,wang2006oxidation}
or the combination of DFT with dynamical mean-field theory
(DMFT).\cite{Kotliar:RMP:2006,held2007electronic,Jacob:PRB:2010a,Karolak:JPCM:2011,Weber:PNAS:2014}
However, these methods introduce additional parameters which can be hard to determine
in an {\it ab initio} fashion and may still fail to properly describe a system in
specific circumstances.\cite{zaki2014failure,huang2016much}

While DFT is usually formulated in real space and applied to real materials and molecules, it is also possible to
formulate a DFT for lattice models, such as the Hubbard and Anderson impurity models.\cite{GunnarssonSchoenhammer:86,SchoenhammerGunnarssonNoack:95,LimaSilvaOliveiraCapelle:03,carrascal2015hubbard,KurthStefanucci:17}
This allows to incorporate electron-electron interactions in a controlled way, and study their effect on the
Hartree-exchange-correlation (Hxc) functional.
Lattice DFT (LDFT) studies have for example revealed the ubiquitous presence of steps in the exact functionals.\cite{SobrinoKurthJacob:20}
It turns out that these steps are crucial for DFT to capture important aspects of strongly correlated systems
by DFT, such as the Kondo plateau in the zero-bias conductance\cite{StefanucciKurth:11,BergfieldLiuBurkeStafford:12,TroesterSchmitteckertEvers:12}
or the opening of a Mott gap.\cite{LimaSilvaOliveiraCapelle:03,KarlssonPriviteraVerdozzi:11}
Furthermore, an extension of DFT which incorporates the current through an interacting system
in addition to its density, called steady-state DFT or i-DFT, is capable of capturing strongly
correlated phenonomena both in the conductance and the many-body spectral function,
such as Coulomb blockade, Kondo effect and the Mott metal-insulator transition.\cite{StefanucciKurth:15,KurthStefanucci:16,JacobKurth:18,SobrinoAgostaKurth:19,JacobStefanucciKurth:20}
Again the proper description of these phenomena is typically
linked to the presence of steps in the corresponding Hxc potentials.

An interesting question then is whether the insights obtained from LDFT can somehow be
exploited to improve the performance of standard functionals of real-space DFT for
molecules and materials in the presence of strong electronic correlations. The challenge
lies in finding features of strong correlations that are generic enough to incorporate in
approximate Hxc functionals in order to comply with the universal character of the true
functional. A good testbed is the dissociation of the hydrogen (H$_2$) molecule, which
despite its apparent simplicity presents a challenge for many electronic structure methods
including DFT. Most approximate functionals encounter difficulties in accurately stretching
H$_2$ without breaking the spin symmetry.\cite{cohen2012challenges}
At its heart H$_2$ dissociation is a strongly correlated problem:
at large bond distances the many-body ground state acquires a pronounced multi-determinant
character which poses a challenge to normal KS-DFT, as explained above.

In this work we first introduce the lattice DFT approach for a two-level system with
different interactions (Sec.~\ref{ldft_two_level}) for which we present the Hxc energy and
potential at $N=2$ and functionals at low temperature. In Sec.~\ref{ldft_rsdft} we propose
a formal connection between lattice and real-space DFT and we identify the interactions in
terms of KS orbitals. In Sec.~\ref{h2}, we apply this connection to the binding energy
curve of the hydrogen molecule in minimal basis and beyond. Finally, we present our
conclusions in Sec.~\ref{conclusions}.

\section{Lattice density functional theory for a two-level system}
\label{ldft_two_level}

First we review some results for the Hxc potentials for the
particular interactions studied in Ref.~\onlinecite{SobrinoKurthJacob:20}. Then we will
introduce an additional term to the interaction, the pair-hopping interaction, which at zero
temperature, can be treated analytically within a density functional framework. 

We consider a two-level system described by the Hamiltonian
\be
\hat{H} = \sum_{i=1}^2 \varepsilon_i \hat{n}_i + \hat{W}
\label{hamil}
\ee
where $\varepsilon_i$ is the single-particle energy of level $i$ and
$\hat{n}_i=\sum_{\sigma=\uparrow,\downarrow} \hat{n}_{i\sigma}$ with
$\hat{n}_{i\sigma}= \hat{c}^{\dagger}_{i\sigma}\hat{c}_{i\sigma}$ and the $\hat{c}^{\dagger}_{i\sigma}$
($\hat{c}_{i\sigma}$) are the creation (annihilation) operators for an electron with spin
$\sigma$ in level $i$. Finally, $\hat{W}$ is the electron-electron interaction whose exact
form will be specified below. We emphasize that the non-interacting part of the Hamiltonian
of Eq.~(\ref{hamil}) is diagonal and that spin symmetry is not broken. In the following we
work in the grand-canonical ensemble at finite temperature $T=\frac{1}{\beta}$ and chemical
potential $\mu$. As usual, expectation values of any operator $\hat{O}$ are given by
$O={\rm Tr}\{\hat{\rho} \hat{O}\}$ where the trace is over a complete set of states spanning
the Fock space and the statistical operator is defined as
\be
\hat{\rho} = \frac{1}{Z} \exp\left(-\beta (\hat{H} - \mu \hat{N}) \right) \;.
\ee
Here, $Z={\rm Tr}\{\hat{\rho}\}$ is the partition function and $\hat{N}=\sum_{i=1}^2 \hat{n}_i$
is the operator for the total electron number.

Of course, since our model is so simple (and the corresponding Fock space small), it is
straightforward to (numerically) exactly diagonalize the Hamiltonian $\hat{H}$ for any
conceivable interaction $\hat{W}$. However, a conceptually alternative approach for the
calculation of the ``densities'' $n_i$ is to use DFT in its
incarnation of LDFT \cite{GunnarssonSchoenhammer:86,LimaSilvaOliveiraCapelle:03}.
Moreover, since we work in the grand-canonical ensemble, the proper DFT framework is given by
Mermin's finite temperature DFT \cite{Mermin:65}. Therefore, we are looking for a
non-interacting Hamiltonian, the Kohn-Sham (KS) Hamiltonian, of the form
\be
\hat{H}_s = \sum_{i=1}^2 \varepsilon_i^s \hat{n}_i
\label{ks_hamil}
\ee
with KS orbital energies $\varepsilon_i^s$ which yields the same densities $n_i$ as the
interacting one. The KS orbital energies can be written as
$\varepsilon_i^s=\varepsilon_i + v_{\rm Hxc,i}$ with 
\be
v_{\rm Hxc,i}(n_1,n_2) = \frac{\partial \Omega_{\rm Hxc}(n_1,n_2)}{\partial n_i}
\label{vhxc_ldft}
\ee
where $\Omega_{\rm Hxc}(n_1,n_2)$ is the Hxc contribution to the grand-canonical potential.
Of course, the exact form of $\Omega_{\rm Hxc}(n_1,n_2)$ depends on the interaction
$\hat{W}$ and, in the zero-temperature limit, it reduces to the Hxc energy contribution
$E_{\rm Hxc}(n_1,n_2)$ to the total ground state energy. 
For given, fixed single-particle energies $\varepsilon_i$ of the
interacting Hamiltonian (\ref{hamil}), the self-consistent KS equations for our model then
read
\be
n_i = 2 f(\varepsilon_i^s) = 2 f(\varepsilon_i + v_{\rm Hxc,i}(n_1,n_2))
\label{ks_eqs}
\ee
where $f(x)= (1 + \exp(-\beta (x-\mu)))^{-1}$ is the Fermi function at inverse temperature
$\beta$ and chemical potential $\mu$ and the prefactor 2 is due to spin degeneracy. Since
the r.h.s. of Eq.~(\ref{ks_eqs}) depends on both densities, the two equations (\ref{ks_eqs})
for $i\in\{1,2\}$ are coupled and have to be solved simultaneously.

In Ref.~\onlinecite{SobrinoKurthJacob:20} we studied a two-level system with interactions
of the form
\bea
\hat{W}_1 &=& \sum_i U_i \, \hat{n}_{i\uparrow} \hat{n}_{i\downarrow}
+ U_{12} \, \hat{n}_{1} \hat{n}_{2} \nn\\
&& - J \left( \sum_{\sigma} \hat{n}_{1\sigma} \hat{n}_{2\sigma} +
\hat{c}^{\dagger}_{1\uparrow}\hat{c}_{1\downarrow}
\hat{c}^{\dagger}_{2\downarrow}\hat{c}_{2\uparrow} +
\hat{c}^{\dagger}_{1\downarrow}\hat{c}_{1\uparrow}
\hat{c}^{\dagger}_{2\uparrow}\hat{c}_{2\downarrow} \right)
\label{inter_wo_pair}
\eea
and found from reverse-engineering of exact results that the corresponding Hxc potentials
in the zero-temperature limit have a structure dominated by step features which are
intimately related to the famous derivative discontinuity of DFT.\cite{PerdewParrLevyBalduz:82}
Even in the zero-temperature limit, the exact form of the Hxc potential depends on the relative
magnitude of the different interaction parameters. For parameters $U_1,U_2 \geq U_{12} \geq J$
we found that the Hxc potential for level $i$ can be decomposed as
\bea
\lefteqn{
v_{\rm Hxc, i}(n_1,n_2) = v_{\rm Hxc}^{\rm CIM}(U_{12}-J)[N]} \nn\\
&& + v_{\rm Hxc}^{\rm SSM}(U_i-U_{12}+J)[n_i] + v_{\rm Hxc}^{\rm Inter}(J/2)[N]
\label{xcpot_reg_I}
\eea
where
\be
v_{\rm Hxc}^{\rm CIM}(U)[N] = \sum_{k=1}^3 U \theta(N-k)
\label{cim}
\ee
is the Hxc potential of the Constant Interaction Model
(CIM)\cite{StefanucciKurth:13,KurthStefanucci:17} with
$N=n_1+n_2$
\be
v_{\rm Hxc}^{\rm SSM}(U)[n_i] = U \theta(n_i-1)
\label{SSM}
\ee
is the Hxc potential of the Single Site Model (SSM) \cite{StefanucciKurth:11}
and
\be
v_{\rm Hxc}^{\rm Inter}(U)[N] =  U \theta(N-2)
\label{inter}
\ee
is the interorbital term. Here we would also like to emphasize that in
Eqs.~(\ref{xcpot_reg_I}) - (\ref{inter}) the function $\theta(x)$
should always be read as a function which in the zero-temperature limit approaches the
Heaviside step function but in contrast to the Heaviside function remains continuous for any
finite temperature $T$. By integrating Eq.~(\ref{xcpot_reg_I}) one can then also derive the
Hxc energy which reads
\bea
\lefteqn{
  E_{\rm Hxc}^{(1)}(n_1,n_2) = (U_{12}-J) \sum_{k=1}^3 (N-k) \theta(N-k)} \nn\\
&& + \sum_{i=1}^2 (U_i-U_{12}+J) (n_i-1) \theta(n_i-1) \nn\\
&& + \frac{J}{2} (N-2) \theta(N-2)
\label{ehxc_reg_I}
\eea
Finally we would like to emphasize a peculiarity of the Hamiltonian (\ref{hamil}) with
interaction $\hat{W}=\hat{W}_1$ of Eq.~(\ref{inter_wo_pair}): for this particular interaction,
all eigenstates of $\hat{H}$ are at the same time eigenstates of the operators
$\hat{n}_i$ with integer eigenvalues. 

\subsection{Including the pair-hopping interaction}
\label{pair_hopping}

Of course, the interaction of Eq.~(\ref{inter_wo_pair}) is not the most general interaction
possible in a two-level system. One generalization additionally includes the so-called
pair hopping interaction (with strength $P$) which is written as
\be
\hat{W}_2 = \hat{W}_1
- P \left( \hat{c}^{\dagger}_{1\uparrow}\hat{c}^{\dagger}_{1\downarrow}
\hat{c}_{2\uparrow}\hat{c}_{2\downarrow} +
\hat{c}^{\dagger}_{2\uparrow}\hat{c}^{\dagger}_{2\downarrow}
\hat{c}_{1\uparrow}\hat{c}_{1\downarrow} \right) \;. 
\label{inter_with_pair}
\ee
We here restrict ourselves to the two-particle sector for which we choose the following
2-electron basis
\bea
  |1 \rangle = \hat{c}_{1 \uparrow }^{\dagger} \hat{c}_{1 \downarrow}^{\dagger} | 0 \rangle
  & \hspace*{7mm} & |4 \rangle
  = \hat{c}_{1 \downarrow}^{\dagger} \hat{c}_{ 2 \uparrow}^{\dagger} | 0\rangle \nn\\
  |2 \rangle = \hat{c}_{1 \uparrow}^{\dagger} \hat{c}_{ 2 \uparrow}^{\dagger}| 0 \rangle &&  |5 \rangle
  = \hat{c}_{1 \downarrow}^{\dagger} \hat{c}_{2 \downarrow}^{\dagger} | 0\rangle \nn\\
  |3 \rangle = \hat{c}_{1 \uparrow}^{\dagger} \hat{c}_{2 \downarrow}^{\dagger} | 0 \rangle &&
  |6 \rangle = \hat{c}_{2 \uparrow}^{\dagger} \hat{c}_{2 \downarrow}^{\dagger} | 0 \rangle
\label{basis}
\eea
where $| 0 \rangle$ is the vacuum state. It is easy to check that states $|2\rangle$ and
$|5\rangle$ are two degenerate eigenstates of the Hamiltonian (\ref{hamil}) (with
$\hat{W}=\hat{W}_2$) with eigenvalue $E_{2/5}=\varepsilon_1 + \varepsilon_2 + U_{12}-J$. From
$|3\rangle$ and $|4\rangle$ we form the two eigenstates
\bea
| \tilde{3} \rangle &=& \frac{1}{\sqrt{2}} (|3\rangle - |4\rangle)
\\
| \tilde{4} \rangle &=& \frac{1}{\sqrt{2}} (|3\rangle + |4\rangle) 
\eea
with eigenvalues
\bea
E_{\tilde{3}} &=& \varepsilon_1 + \varepsilon_2 + U_{12}+J \nn\\
E_{\tilde{4}} &=& \varepsilon_1 + \varepsilon_2 + U_{12}-J \; ,
\eea
i.e., $| \tilde{4} \rangle$ is degenerate with $|2\rangle$ and $|5\rangle$
(spin triplet\st{s}). The remaining eigenstates $| \tilde{1} \rangle$ and
$| \tilde{6} \rangle$ can be written as linear combination of states $| 1 \rangle$ and
$| 6 \rangle$
\bea
| \tilde{1} \rangle &=& \cos \vartheta |1\rangle + \sin \vartheta |6\rangle
\label{groundstate}\\
| \tilde{6} \rangle &=& -\sin \vartheta |1\rangle + \cos \vartheta |6\rangle
\label{state_deln}
\eea
with some parameter $\vartheta$ to be specified and with eigenvalues
\bea
E_{\tilde{1}/\tilde{6}} &=& \varepsilon_1 + \varepsilon_2 + \frac{1}{2} (U_1+U_2) \nn\\
&& \mp \frac{1}{2}\sqrt{(2 \varepsilon_1 - 2 \varepsilon_2 + U_1 - U_2)^2 + 4 P^2} \;.
\label{e_gs}
\eea
We note that out of all eigenstates, only states $| \tilde{1} \rangle $ and
$| \tilde{6} \rangle $ can lead to a non-vansihing density difference $\Delta n = n_1-n_2$.
Without loss of generality (since $\vartheta$ hasn't been specified yet), we assume that out
of the two eigenstates $| \tilde{1} \rangle $ and $| \tilde{6} \rangle $,
$| \tilde{1} \rangle $ is the one with lower eigenvalue. The densities of this state can
easily be computed as 
\bea
n_1 &=& \langle \tilde{1} | \hat{n}_1 | \tilde{1} \rangle = 2 \cos^2 \vartheta
\nn\\
n_2 &=& \langle \tilde{1} | \hat{n}_2 | \tilde{1} \rangle = 2 \sin^2 \vartheta
\label{dens_angle}
\eea
or
\be
\Delta n = n_1-n_2 = 2 (\cos^2 \vartheta - \sin^2 \vartheta) \;.
\ee
Eqs.~(\ref{dens_angle}) can then be used to express
\bea
\cos^2 \vartheta &=& \frac{1}{2} \left(1 + \frac{\Delta n}{2}\right) \nn\\
\sin^2 \vartheta &=& \frac{1}{2} \left(1 - \frac{\Delta n}{2}\right)
\eea
For our model, the Hohenberg-Kohn (HK) functional (which only depends on
$\Delta n$) can be found from the constrained search approach as
\be
F(\Delta n) = \min_{\Psi \to \Delta n} \langle \Psi | \hat{W}_2 | \Psi \rangle \;.
\ee
Since for given $\Delta n \neq 0$, the minimizing state $|\Psi_0\rangle$ must have the form
(\ref{groundstate}), the HK functional can easily be evaluated as 
\bea
\lefteqn{
F(\Delta n) = \langle \tilde{1} | \hat{W}_2 | \tilde{1} \rangle }\nn\\
&& = \frac{U_1}{2} \left(1 + \frac{\Delta n}{2} \right) +
\frac{U_2}{2} \left(1 - \frac{\Delta n}{2} \right)
- P \sqrt{1 - \frac{(\Delta n)^2}{4} } 
\label{hk_func}
\eea
where the minus sign has to be chosen for the last (square root) term for the functional
to be minimal for a given $\Delta n$. For $\Delta n=0$, $F(\Delta n)$ of Eq.~(\ref{hk_func}) 
coincides with the HK functional if the interaction parameters are such that
$(U_1+U_2)/2 -P < U_{12} -J$. For $(U_1+U_2)/2 -P = U_{12} -J$ the states $|\tilde{1}\rangle$
and $|\tilde{4}\rangle$ become degenerate ground states (at $\Delta n=0$) while for
$(U_1+U_2)/2 -P > U_{12} -J$ the state $|\tilde{4}\rangle$ becomes the unique ground state.
It is important to note that the states $|\tilde{1}\rangle$ and
$|\tilde{4}\rangle$ have a very different character. While the former is a spin singlet
formed as linear combination of Slater determinants each with a single orbital occupied
by two electrons, the latter is a triplet formed as linear combination of Slater
determinants each with two different orbitals occupied by a single electron. It is
exactly for this reason that in the limit $J=P=0$, Eq.~(\ref{hk_func}) reduces to
$(U_1+U_2)/2$ (for $\Delta n = 0$) while at the same densities the proper HK functional
for $(U_1+U_2)/2 > U_{12}$ (which follows from integrating Eq.~(13) of
Ref.~\onlinecite{SobrinoKurthJacob:20}) gives the value $U_{12}$ .

Since in our model there is no kinetic energy
contribution (neither in the interacting nor the non-interacting case), the
functional for the Hxc energy is identical to the HK functional
\be
E_{\rm Hxc}(\Delta n) = F(\Delta n)
\label{ehxc_pair}
\ee
and the Hxc potentials are
\bea
v_{\rm Hxc,1}(\Delta n) &=& \frac{U_1-U_2}{4} + \frac{P \Delta n}{4 \sqrt{1 - \frac{(\Delta n)^2}{4}}}
\nn\\
v_{\rm Hxc,2}(\Delta n) &=& - \frac{U_1-U_2}{4}
-\frac{P \Delta n}{4 \sqrt{1 - \frac{(\Delta n)^2}{4}}}
\label{hxc_pot}
\eea
or
\bea
\Delta v_{\rm Hxc}(\Delta n) &=& v_{\rm Hxc,1}(\Delta n) - v_{\rm Hxc,2}(\Delta n)
\nn\\
&=& \frac{U_1-U_2}{2} + \frac{P \Delta n}{2 \sqrt{1 - \frac{(\Delta n)^2}{4}}} \;. 
\eea
Knowing the HK functional $F(\Delta n)$, we can for given values of the single-particle
parameters (on-site energies) $\varepsilon_1 = \frac{\Delta \varepsilon_0}{2}$ and
$\varepsilon_2 = -\frac{\Delta \varepsilon_0}{2}$ find the corresponding ground state
density $\Delta n_0$ by minimizing the HK total energy functional 
\be
E_{\Delta \varepsilon_0}(\Delta n) = \frac{\Delta \varepsilon_0}{2} \Delta n + F(\Delta n) \;,
\label{hk_en_func}
\ee
i.e., we solve
\bea
\lefteqn{
\frac{\partial E_{\Delta \varepsilon_0}(\Delta n)}{\partial \Delta n}
\bigg\vert_{\Delta n=\Delta n_0} = 0 }\nn\\
&=& \Delta \varepsilon_0 + \frac{U_1-U_2}{4} 
+ \frac{P}{4} \frac{\Delta n_0}{\sqrt{1 - \frac{(\Delta n_0)^2}{4}}} \;.
\eea
This can be rearranged for $\Delta n_0$ and gives
\be
\Delta n_0 = - \frac{2 (2 \Delta \varepsilon_0 + U_1 - U_2)}{\sqrt{
    (2 \Delta \varepsilon_0 + U_1 - U_2)^2 + 4 P^2}}
\label{gs_dens}
\ee
where the minus sign was chosen for the physical reason that we should have
$\Delta n_0 \to 2$ for $\Delta \varepsilon_0 \to -\infty$. Reinserting
this density into the HK energy functional (\ref{hk_en_func}), one can easily
confirm that one obtains the ground state energy $E_{\tilde{1}}$ of
Eq.~(\ref{e_gs}).

\begin{figure}
  \includegraphics[width=0.47\textwidth]{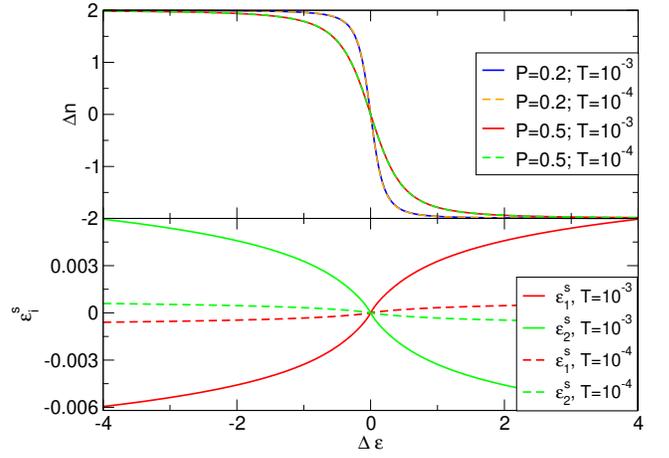}
  \caption{Upper panel: Density imbalance $\Delta n=n_1-n_2$ versus single particle energy
    difference $\Delta \varepsilon = \varepsilon_1-\varepsilon_2$ for the model described by
    the Hamiltonian (\ref{hamil}) with interaction $\hat{W}_2$ of Eq.~(\ref{inter_with_pair})
    for different values of the pair-hopping interaction $P$ and different temperatures $T$.
    The other parameters are $U_1=U_2=U_{12}=1.0$ and $J=P$. Lower panel: KS eigenvalues
    $\varepsilon_i^s = \varepsilon_i + v_{\rm Hxc,i}$ at self-consistency for $P=J=0.5$ and
    two different temperatures $T$. The other parameters are $U_1=U_2=U_{12}=1.0$. All
    energies given in units of $U_1$.
  }
  \label{fig:deldens}
\end{figure}

Above we have computed the ground state density $\Delta n_0$ directly from the
HK variational principle. The same result should, of course, also be obtained
by self-consistent solution of the KS equations (\ref{ks_eqs}). For given
single-particle parameters $\varepsilon_i$, these equations can be conveniently
rewritten as
\be
\Delta n = 2 \left[ f(\varepsilon_1 + v_{\rm Hxc,1}(\Delta n)) - f(\varepsilon_2 +
v_{\rm Hxc,2}(\Delta n)) \right] 
\label{ks_eqs_2}
\ee
and the solution of this equation has to coincide with the ground state density of
Eq.~(\ref{gs_dens}) which we have confirmed numerically. In the upper panel of
Fig.~\ref{fig:deldens} we show the density imbalance $\Delta n$ as function of
$\Delta \varepsilon = \varepsilon_1 - \varepsilon_2$ for different values of the pair hopping
parameter $P$ and different temperatures. The other parameters are $U_1=U_2=U_{12}=1.0$ and $J=P$
and where chosen to ensure that the ground state has the form of Eq.~(\ref{groundstate}) for
any $\Delta \varepsilon$. As expected, the stronger the pair-hopping interaction, the more the
step in $\Delta n$ at $\Delta n=0$ is smoothened. We have solved both the many-body problem
as well as the self-consistent solution of the corresponding KS system and both solutions are
identical on the scale of the figure (not shown). Moreover, we have computed $\Delta n$ for
two different, small temperatures $T$ and found no appreciable difference in the corresponding
density imbalances. However, we would like to draw attention to the particular way in which the
KS system achieves the same density as the interacting one. We first note that the Fermi functions
in the zero temperature limit approaches a step function and naively, Eq.~(\ref{ks_eqs_2}) seems
to suggest that its solution can only be $\pm 2$ which is clearly not the case
for arbitrary values of $\Delta \varepsilon$. Therefore, the Hxc potential must
lead to total KS energies to be close to the chemical potential, i.e., the
step region of the Fermi functions. This is confirmed by looking at the lower panel of
Fig.~\ref{fig:deldens} where we show the self-consistent KS eigenvalues for the parameters of the
upper panel with $P=0.5$ at two small temperatures. While the density already is basically
converged at $T=10^{-3}$, the KS eigenvalues are not. In fact, in the $T\to 0$ limit the KS eigenvalues
will converge to the chemical potential $\mu$ (here taken to be zero) because they have to reproduce the
densities. This can only be achieved by moving the KS eigenvalues onto the step of the Fermi function
whose width is of the order of $T$. Even in this limit, however, the upper and lower KS eigenvalues have
to converge to slightly different values in order to give a finite density imbalance $\Delta n$. This 
highlights the importance of working at small but finite temperature for which the Fermi functions
exhibit a step but are not mathematically discontinuous. This is somewhat related to findings in
previous works \cite{StefanucciKurth:11,KurthStefanucci:13} where it was found that Hxc potentials
which exhibit step stuctures typically lead to a pinning of KS energy levels to the Fermi energy
for a range of single-particle energies.

\section{Establishing a connection between lattice DFT and real-space DFT}
\label{ldft_rsdft}

In the present Section and based on the results of the previous Section, we will establish a 
connection between lattice DFT and the usual DFT formulation in real space. The combination of
standard DFT with advanced many-body tretaments of lattice models has a long tradition, possibly
starting with what is known as DFT+U method \cite{anisimov1991band}. A more recent approach combines dynamical
mean-field theory (DMFT) with DFT \cite{Kotliar:RMP:2006,Jacob:PRB:2010a,Weber:PNAS:2014}. In a
somewhat different line, a direct transfer of ideas from lattice DFT to standard quantum chemical
methods has been suggested recently\cite{Coe:19}. Here we suggest %another possibility of establishing a
an alternative connection between lattice DFT and real-space DFT based on our results for the
two-level system described in Sec.~\ref{ldft_two_level}.

In Mermin's version of finite-temperature DFT \cite{Mermin:65}, the electronic density (in real space)
in thermal equilibrium for an interacting many-electron system subject to an electrostatic potential
$v_0(\vr)$ is determined by self-consistent solution of the KS equation
\be
\left( -\frac{\nabla^2}{2} + v_0(\vr) + v_{\rm Hxc}[\rho](\vr) \right) \varphi_{i \sigma}(\vr)
= \varepsilon_{i \sigma}^{s} \varphi_{i \sigma}(\vr)
\label{ks_eq_mermin}
\ee
with the KS orbitals $\varphi_{i \sigma}$ and the KS energy eigenvalues $\varepsilon_{i \sigma}^{s}$.
The electronic density is given by 
\be
\rho(\vr) = \sum_{i \sigma} n_{i\sigma} |\varphi_{i \sigma}(\vr)|^2
\label{dens_mermin}
\ee
where the occupation of the KS orbital $\varphi_{i \sigma}$ is given by
$n_{i \sigma} = f(\varepsilon_{i \sigma}^{s})$. In Eq.~(\ref{ks_eq_mermin}),
the Hxc potential is given as functional derivative of the Hxc contribution $\Omega_{\rm Hxc}$ to the
grand canonical potential, i.e.,
\be
v_{\rm Hxc}[\rho](\vr) = \frac{\Omega_{\rm Hxc}[\rho]}{\delta \rho(\vr)} 
\label{vhxc_mermin}
\ee
which, of course, has to be approximated in practice.

In Sec.~\ref{ldft_two_level} we have found the Hxc energy $E_{\rm Hxc}$ (i.e., the zero-temperature
limit of $\Omega_{\rm Hxc}$) for the two-level system as function of the occupation numbers $n_i$ and the
interaction parameters $U_i$, $U_{12}$, $J$, and $P$. In order to make a connection between
these lattice DFT results and real-space DFT, we now propose to interpret the occupation
numbers as $n_i=\sum_{\sigma} n_{i \sigma} = \sum_{\sigma} f(\varepsilon_{i \sigma}^{s})$ and the
interaction parameters as two-electron Coulomb integrals with respect to the KS orbitals. We
define a general two-electron integral as
\be
( i\sigma j \sigma | k \sigma' l \sigma') = \Id{r} \Id{r'}
\frac{\varphi_{i\sigma}^*(\vr) \varphi_{j\sigma}(\vr) \varphi_{k\sigma'}^*(\vr') \varphi_{l\sigma'}(\vr') }
     {|\vr - \vr'|}
\label{two_el_int}
\ee
Then the interaction parameters can be identified in terms of these two-electron integrals as
\bea
U_i = (i \sigma i \sigma | i \bar{\sigma} i \bar{\sigma}) & \hspace*{5mm} &
J = (1 \sigma 2 \sigma | 2 \sigma 1 \sigma) \nn\\
U_{12} = (1 \sigma 1 \sigma | 2 \sigma 2 \sigma) & \hspace*{1cm} &
P = (1 \sigma 2 \bar{\sigma} | 1 \bar{\sigma} 2 \sigma)
\label{interactions}
\eea
where $\bar{\sigma}=\downarrow (\uparrow)$ for $\sigma=\uparrow (\downarrow)$. We note that for
real and spin-independent KS orbitals we have $J=P$. With this interpretation, the interaction
parameters formally become functionals of the KS orbitals and the occupation numbers functionals
of the KS energy eigenvalues. In the context of DFT, functionals depending on KS orbitals and
KS eigenvalues are {\em implicit} functionals of the density and it is well known that the
corresponding (Hxc) potentials (i.e., functional derivatives with respect to the density) can be
computed with the Optimized Effective Potential (OEP) formalism
\cite{TalmanShadwick:76,GraboKreibichKurthGross:00,KuemmelKronik:08}.

We now take a look at what our interpretation of the lattice DFT functionals in terms of
real-space KS orbitals and occupation numbers implies for the specific example of $E_{\rm Hxc}^{(1)}$
given by Eq.~(\ref{ehxc_reg_I}). We will not consider general values for $n_1$ and $n_2$ but
instead focus on those points where both occupation numbers are integer. For $n_1=n_2=0$ as well as
for $n_1=1$, $n_2=0$ and $n_1=0$, $n_2=1$ we see that $E_{\rm Hxc}^{(1)}$ vanishes,
i.e., $E_{\rm Hxc}^{(1)}$ is self-interaction free for one electron. For $n_1=2$ and
$n_2=0$ we find $E_{\rm Hxc}^{(1)}(n_1=2,n_2=0) = U_1$ which can be written as
\be
E_{\rm Hxc}^{(1)}(n_1=2,n_2=0) = \frac{1}{4} \Id{r} \Id{r'} \frac{ \rho(\vr) \rho(\vr')}{| \vr - \vr'|}
\label{ehxc_reg_1_N2}
\ee
where we assumed spin-independent KS orbitals such that the density becomes
$\rho(\vr) = 2 |\varphi_{1 \sigma}(\vr)|^2$. The same result (in terms of the density) is obtained
for $n_1=0$, $n_2=2$. For full occupation of the two-level system, $n_1=2$, $n_2=2$, we obtain
$E_{\rm Hxc}^{(1)}(n_1=2,n_2=2) = 4 U_{12} + U_1 +U_2 - 2J$  which can be written as 
\bea
\lefteqn{
  E_{\rm Hxc}^{(1)}(n_1=2,n_2=2) = \frac{1}{2} \Id{r} \Id{r'} \frac{\rho(\vr) \rho(\vr')}{| \vr - \vr'|} }
\nn\\
&& - \frac{1}{2} \sum_{\sigma} \sum_{i,j=1}^2 \Id{r} \Id{r'}
\frac{\varphi_{i \sigma}^*(\vr) \varphi_{j \sigma}(\vr)
  \varphi_{2 \sigma}^*(\vr') \varphi_{1 \sigma}(\vr')}{| \vr - \vr'|} \;. 
\label{ehxc_reg_1_N4}
\eea
We recognize that both Eq.~(\ref{ehxc_reg_1_N2}) and (\ref{ehxc_reg_1_N4}) are nothing but the
Hartree plus the exact exchange energy of standard DFT. In fact, also for the other integer
occupations the functional of Eq.~(\ref{ehxc_reg_I}) can with our interpretation (lattice densities identified
as occupations of KS orbitals and interaction parameters defined in terms of KS orbitals) be
identified as the Hartree plus exact exchange functionals. 
However, for occupations $n_1=2$, $n_2=1$ and $n_1=1$, $n_2=2$ as well as for $n_1=n_2=1$ this identification
requires the use of the proper definition of the Hartree plus exact exchange energy of
ensemble DFT derived in Refs.~\onlinecite{GouldPittalis:17,GouldPittalis:19}.
Based on the recovery of the Hartree plus exact exchange energy for integer
occupations, we may infer that the energy functional of Eq.~(\ref{ehxc_reg_I}) is actually the
proper generalization of this functional for any non-integer occupation $0\leq n_i \leq 2$.
Furthermore, from the recovery of a known functional from Eq.~(\ref{ehxc_reg_I}), we also gain
confidence that the functional of Eq.~(\ref{ehxc_pair}) with our interpretation is
%at least in some sense
a reasonable approximation to the exact Hxc functional of a two-level system. This
will be borne out in the next Section where we apply the functional (\ref{ehxc_pair}) to the
description of the ${\rm H}_2$ molecule.

\section{Application to the hydrogen molecule}
\label{h2}

In the present Section we describe an application of the formalism presented in
Sec.~\ref{ldft_two_level} to the hydrogen molecule. We start with the problem treated
with a minimal basis set before investigating larger basis sets.

\subsection{Minimal basis set}
\label{min_basis}

In the minimal basis for the hydrogen molecule (H$_2$), for each atom we have only one
single $s$-type basis function. Let $g(\vr)(=g(-\vr))$ be such an $s$-type normalized
(real) basis function localized at the origin $\vr_0={\bf 0}$. If we take the hydrogen
atoms to be located at $\pm \vR/2$, from the localized basis functions
$g_{1/2}(\vr) = g(\vr \pm \vR/2)$ we can construct two normalized and orthogonal (spin)
orbitals, one bonding and one anti-bonding, which take the form 
\bea
\varphi_1(\vr) &=& \varphi_{1 \sigma}(\vr) = 
\frac{1}{\sqrt{2 ( 1 + S)}} \left( g_1(\vr) + g_2(\vr) \right) \nn \\
\varphi_2(\vr) &=& \varphi_{2 \sigma}(\vr) = 
\frac{1}{\sqrt{2 ( 1 - S)}} \left( g_1(\vr) - g_2(\vr) \right)
\label{orbitals_minbas}
\eea
where
\be
S = \Id{r} \; g_1(\vr) g_2(\vr)
\ee
is the overlap integral of the two localized basis functions. We note that by construction
the $\varphi_k(\vr)$ are eigenfunctions of the parity operator, i.e., they satisfy the symmetry
relations
\be
\varphi_1(-\vr) = \varphi_1(\vr) \hspace*{1cm}
\varphi_2(-\vr) = -\varphi_2(\vr)
\label{parity}
\ee
When written in terms of field operators, the Hamiltonian of the hydrogen molecule is given by
\bea
\lefteqn{
  \hat{H}_{\rm H_2} = \sum_{\sigma} \Id{r} \; \hat{\psi}_{\sigma}^{\dagger}(\vr) \hat{h}_0(\vr)
  \hat{\psi}_{\sigma}(\vr) + E_{\rm nuc}(R)}\nn\\
&& \!\!\! + \frac{1}{2} \sum_{\sigma,\sigma'} \Id{r} \Id{r}' \; \hat{\psi}_{\sigma}^{\dagger}(\vr)
\hat{\psi}_{\sigma'}^{\dagger}(\vr') \frac{1}{| \vr - \vr' |}
\hat{\psi}_{\sigma'}(\vr') \hat{\psi}_{\sigma}(\vr) 
\label{hamil_h2_real_space}
\eea
with the single-particle Hamiltonian 
\be
\hat{h}_0(\vr) = -\frac{\nabla^2}{2} + v(\vr) 
\label{sp_hamil}
\ee
where $v(\vr)$ is the attractive potential due to the protons  
\be
v(\vr) = - \frac{1}{|\vr - \vR/2|} - \frac{1}{|\vr + \vR/2|} \;,
\ee
and the nuclear electrostatic repulsion energy $E_{\rm nuc}(R) = 1/R$ with $R=|\vR|$.

In the minimal basis, the field operators can be written as
\be
\hat{\psi}_{\sigma}^{\dagger}(\vr) = \sum_{i=1}^2 \varphi_{i \sigma}^{\ast}(\vr)
\hat{c}_{i \sigma}^{\dagger} 
\ee
where the $\hat{c}_{i \sigma}^{\dagger}$ are the creation operators for an electron in
orbital $\varphi_{i\sigma}$. When inserted into Eq.~(\ref{hamil_h2_real_space}) the
Hamiltonian of the H$_2$ molecule
in the minimal basis (mb) takes the form
\bea
\hat{H}_{\rm H_2}^{\rm mb} &=& \sum_{i=1}^2 \varepsilon_i \hat{n}_i  + E_{\rm nuc}(R) \nn\\
&& + \frac{1}{2} \sum_{\sigma,\sigma'} \sum_{ij,k,l=1}^2 ( i\sigma j \sigma' | k \sigma' l \sigma)
\hat{c}_{i\sigma}^{\dagger} \hat{c}_{j\sigma'}^{\dagger} \hat{c}_{k\sigma'} \hat{c}_{l\sigma}
\label{hamil_h2_mb}
\eea
where the single-particle energies are given by
\be
\varepsilon_{i} = \langle i \sigma | \hat{h}_0(\vr) | i \sigma \rangle
\label{mat_el_core}
\ee
and the off-diagonal matrix elements of $\hat{h}_0(\vr)$ vanish due to the symmetry
$\hat{h}_0(-\vr)=\hat{h}_0(\vr)$ together with the symmetry (\ref{parity}) of the
basis functions, i.e., 
\be
\langle i \sigma | \hat{h}_0(\vr) | j \sigma' \rangle = \delta_{\sigma \sigma'} \delta_{i j}
\Id{r} \; \varphi_{i \sigma}^*(\vr) \hat{h}_0(\vr) \varphi_{i \sigma}(\vr) =
\delta_{\sigma \sigma'} \delta_{i j} \varepsilon_i \;.
\label{core_energies}
\ee
Again, due to the symmetry (\ref{parity}) of the basis functions, it is easy to show that all
those two-electron integrals vanish for which three of the four indices  $\{i,j,k,l\}$ are equal 
and the Hamiltonian for H$_2$ in the minimal basis becomes
\be
\hat{H}_{\rm H_2}^{\rm mb} = \sum_{i=1}^2 \varepsilon_i \hat{n}_i + \hat{W}_2 + E_{\rm nuc}(R)\;. 
\ee
Since this Hamiltonian now has exactly the form of the one studied in the
Sec.~\ref{ldft_two_level} (plus the additive nuclear repulsion energy), the total energy
functional of H$_2$ takes the form
\be
E_{\rm H_2}(\Delta n) = \sum_{i=1}^2 \varepsilon_i n_i + F(\Delta n) + E_{\rm nuc}(R)
\label{etot_2lev}
\ee
with the HK functional of Eq.~(\ref{hk_en_func}). The direct minimization 
for $\Delta n$ as well as the
solution of the corresponding KS equation (\ref{ks_eqs_2}) proceed as described in
Sec.~\ref{ldft_two_level} and the resulting difference in occupation numbers is given by
Eq.~(\ref{gs_dens}). 
Note that here we are only minimizing the occupation number difference
while the orbitals of Eq.~(\ref{orbitals_minbas}) remain fixed. 
\begin{figure}
  \includegraphics[width=0.47\textwidth]{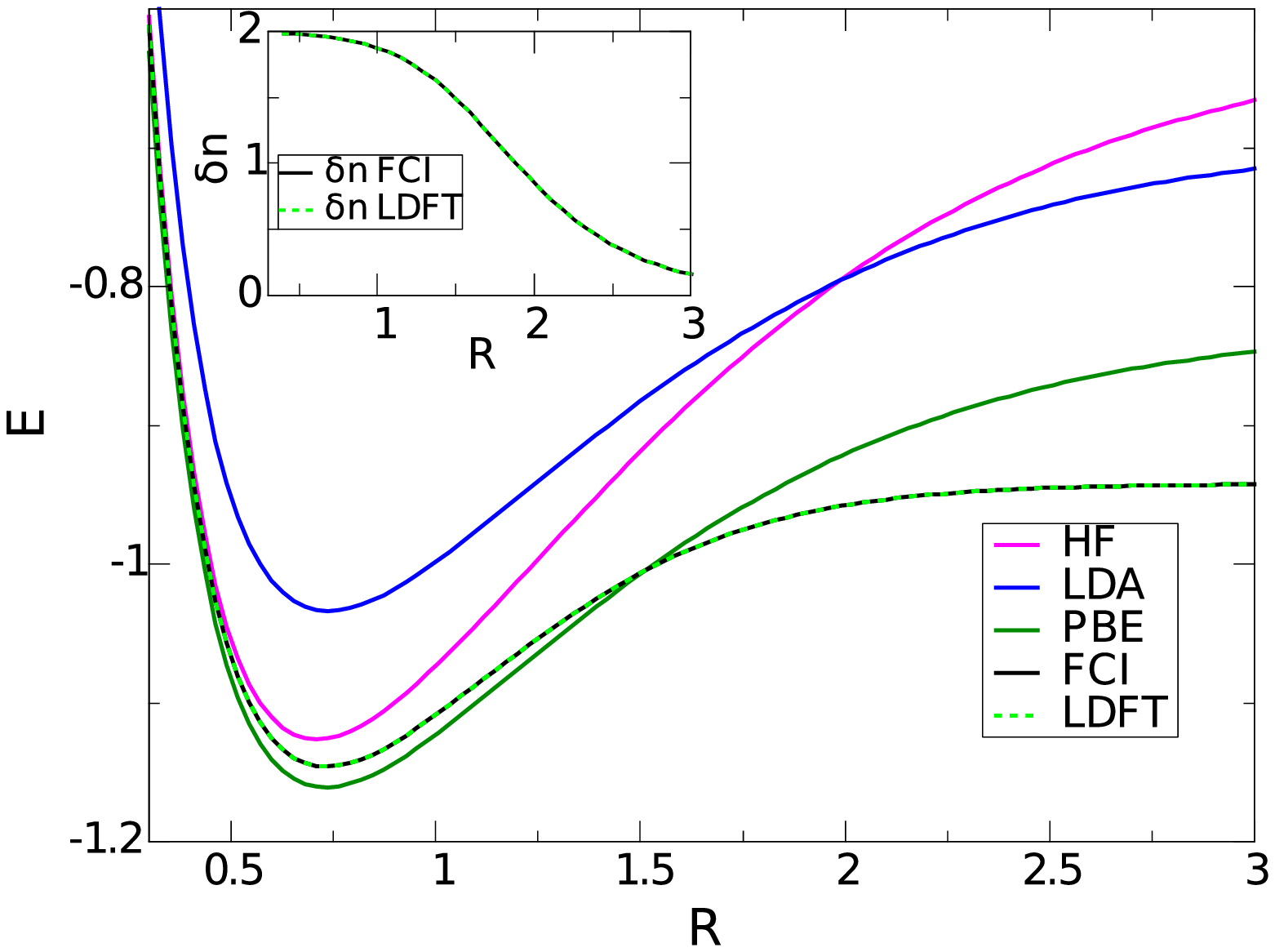}
  \caption{Binding energy of the hydrogen molecule in minimal basis as function of the internuclear
    separation $R$ for various spin-restricted approaches: 
    Hartree-Fock (HF) as well as the DFT approaches using the LDA and PBE functionals and our
    functional (LDFT). The full CI result is given as reference. In the inset we show the
    occupation number difference in full CI as function of $R$ which is identical to our LDFT result.
    All quantities given in atomic units.
  }
  \label{fig:etot_minbas}
\end{figure}

We have mapped out the binding energy curve for the H$_2$ molecule in 
minimal basis with the open source PySCF code\cite{PySCF}. We calculated, for each
internuclear distance $R$, the interaction
parameters $U_i$ and $P$ as well as the corresponding matrix elements (\ref{mat_el_core})
which enter in the evaluation of the occupation numbers (Eq.~(\ref{gs_dens})).
The total energy was then computed by adding the
internuclear repulsion to the total energy functional (\ref{hk_en_func}). In the minimal
basis, our approach becomes exact and therefore equivalent to full configuration interaction
(FCI) which can be confirmed analytically \cite{Szabo:book:1989}.
In Fig.~\ref{fig:etot_minbas} we show the
binding energy curves from our approach and compare with standard spin-restricted Hartree-Fock (HF)
as well as DFT calculations using the LDA and PBE functionals\cite{PerdewBurkeErnzerhof:96}. As
expected, our approach recovers the full CI results while both spin-restricted HF and standard DFT,
as is well-known, do not recover the correct large separation limit (see, however,
the partition DFT of Refs.~\onlinecite{NafzigerWasserman:15,ShiShiWasserman:23} which also
captures dissociation without breaking the spin symmetry).
The reason why our approach does indeed recover this limit simply lies in the
fact that both the ground-state Slater determinant
as well as the doubly excited Slater determinant contribute, i.e., the occupation numbers $n_1$ and
$n_2$ are not strictly integer. In the context of our DFT approach this is possible because we are
working with the equilibrium grand-canonical {\em ensemble} of non-interacting KS wavefunctions and
not just with the KS ground state wavefunction.

\subsection{Beyond the minimal basis}
\label{beyond_min}

\begin{figure}
  \includegraphics[width=0.47\textwidth]{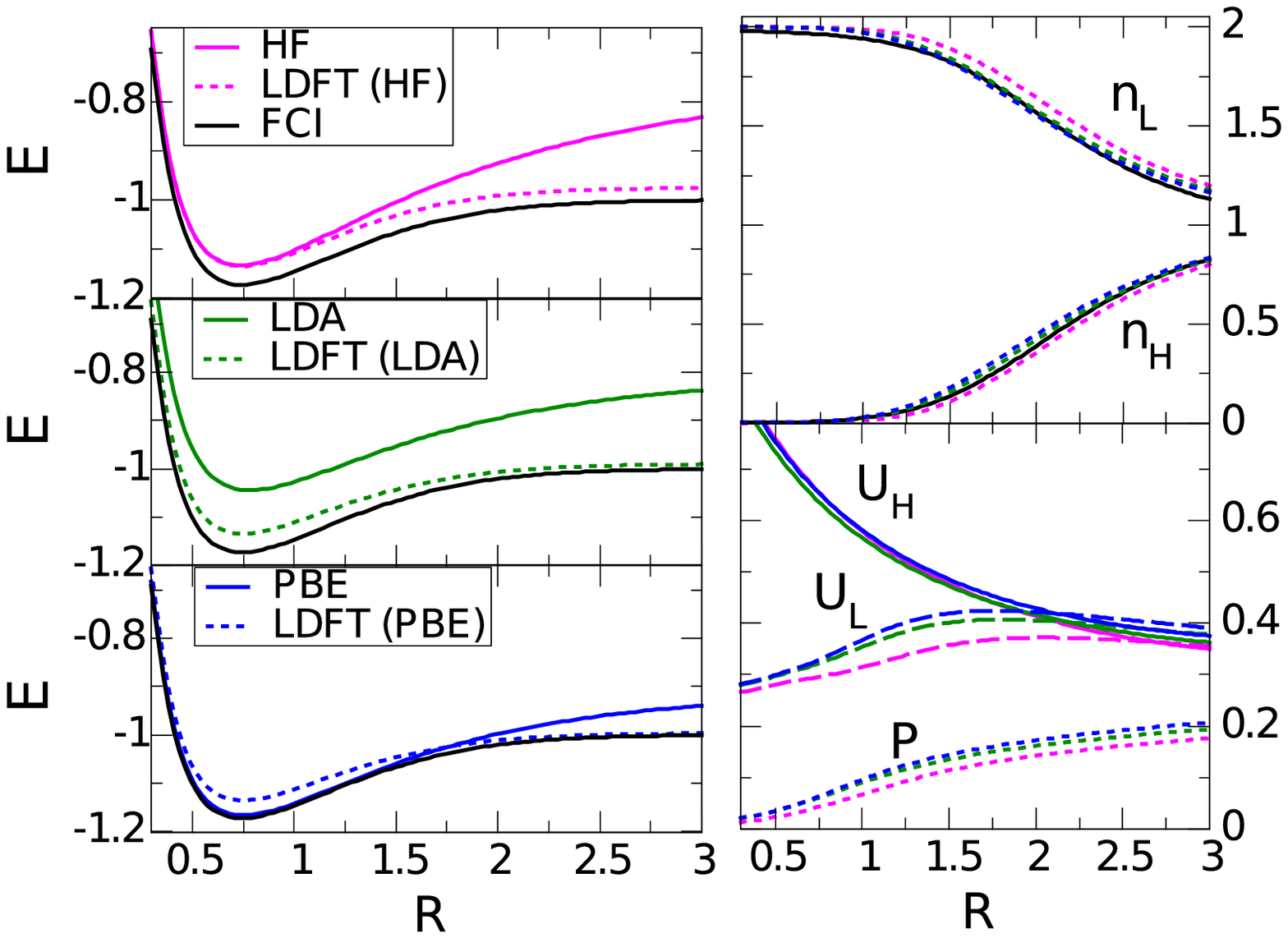}
  \caption{Left panel: Binding energy of the hydrogen molecule in cc-pvtz basis as function of
    internuclear separation $R$ using HF, LDA, and PBE
    (spin-restricted) self-consistent orbitals to evaluate the energy with our approach in comparison
    to full CI results. Upper right panel: occupation numbers of the HOMO and LUMO KS orbitals as
    function of $R$ computed from different self-consistent orbitals. Lower right panel: interaction
    parameters as function of $R$ from different orbitals. All quantities given in atomic units.
  }
  \label{fig:etot_occ_ccpvtz}
\end{figure}

We have also evaluated the binding energy curves of the hydrogen molecule within our approach but
using a larger basis. To this end, we performed self-consistent HF, LDA, and PBE calculations
with the larger cc-pvtz basis\cite{PySCF}. From the resulting
lowest-lying molecular KS orbitals, the HOMO
orbital $\varphi_{\rm H}(\vr) = \varphi_{1 \sigma}(\vr)$ and orbital energy
$\varepsilon_{\rm H} = \varepsilon_1$ as well as the LUMO orbital
$\varphi_{\rm L}(\vr) = \varphi_{2 \sigma}(\vr)$ with orbital energy 
$\varepsilon_{\rm L} = \varepsilon_2$ we evaluate both the single-particle matrix elements
of Eq.~(\ref{mat_el_core}) as well as all the interaction parameters.
Note that the orbital energies $\varepsilon_i$ are \emph{not} KS energy
eigenvalues but the expectation values of the \emph{core} Hamiltonian in the KS orbitals, as
defined by Eq.~(\ref{mat_el_core}). Once these parameters
are determined, we evaluate the total energy according to Eq.~(\ref{etot_2lev}),
i.e., we find the orbital occupations $n_i$ which minimize the total energy
according to our two-level model while keeping the orbitals fixed.

In the left panel of Fig.~\ref{fig:etot_occ_ccpvtz} we show the binding energy curves obtained
in our approach from the three different sets of self-consistent molecular (spin-restricted)
orbitals obtained with the three different functionals in comparison to the corresponding
standard spin-restricted and full CI results. As a common theme, for the three sets of
orbitals the large separation limit is captured correctly with our approach
unlike for the spin-restricted self-consistent calculations. The closest results to the
full CI reference in this limit is achieved with PBE, followed by LDA and HF. Around the
equilibrium bond distance the three sets of orbitals lead to similar results, very close to
self-consistent HF. LDA orbitals give significantly improved total energies as
compared to self-consistent LDA results,
while PBE orbitals give slightly worse energies than self-consistent PBE ones.

In the upper right panel of Fig.~\ref{fig:etot_occ_ccpvtz} we compare orbital occupations from our
approach with the three sets of orbitals. They all follow reasonably well the exact reference results
with HF performing slightly worse. The interaction parameters (lower right panel of
Fig.~\ref{fig:etot_occ_ccpvtz}) from the three sets of orbitals are reasonably similar among each
other, again with HF showing more pronounced differences to the DFT results, especially for the
interaction parameter $U_{\rm L}$ corresponding to the LUMO orbital. 

As discussed above, the results of the present Section for our approach were not computed
self-consistently but with orbitals obtained either from HF or other DFT functionals. In
principle our two-level functional can be read as an orbital-dependent functional which
requires the OEP method for calculation of the corresponding self-consistent potential which,
however, is beyond the scope of the present work. 

\section{Summary and conclusions}
\label{conclusions}

The present work was motivated by earlier work \cite{SobrinoKurthJacob:20} on a simple
model of a double quantum dot on a lattice subject to different types of electron-electron
interactions. We aimed at finding possible connections between lattice DFT
(here in the framework of Mermin's finite-temperature version of DFT in the grand-canonical
ensemble\cite{Mermin:65}) and real-space DFT. This connection could be found if the
interaction paramaeters of the lattice DFT model are read as two-electron Coulomb integrals 
with respect to the KS orbitals of real-space DFT and the lattice ``densities'' are
interpreted as the occupation numbers of these KS orbitals. For the interactions studied in
Ref.~\onlinecite{SobrinoKurthJacob:20} we found that, for integer occupation of the orbitals,
this interpretation leads to a recovery of the exact-exchange energy functional.

We also studied an additional term to the interaction in the lattice model, the pair-hopping
interaction, for which we found the analytical form of the exact HK energy functional. As a
crucial difference to the previously studied interactions, we found that the pair-hopping
term leads to non-integer occupations of the KS orbitals even in the limit of zero
temperature. Another connection to real-space DFT could be established by showing that the
Hamiltonian of the hydrogen molecule (H$_2$), when treated in a minimial basis of localized
basis functions, has exactly the form of our lattice model with the pair-hopping term
included. Knowing the exact HK energy functional (and therefore also the exact Hxc energy
functional) for this model, the binding energy curve of H$_2$ was found to coincide with
the full CI one in the minimal basis for {\em all} internuclear distances and without the
need of breaking spin symmetry. This is achieved by both the KS ground state and the doubly
excited KS determinant having finite weight in the KS ensemble (in the zero-temperature
limit) which is equivalent to saying that both HOMO and LUMO KS orbitals have non-integer
occupation in this ensemble. Finally, we suggested a post-SCF evaluation of our energy
functional for larger basis sets allowing to recover the correct large-separation limit
without spin symmetry breaking. This latter approach still requires a self-consistent
treatment, e.g., within the OEP approach, which will be the subject of future work.

\begin{acknowledgments}
  We acknowledge financial support through Grant PID2020-112811GB-I00 funded by
  MCIN/AEI/10.13039/501100011033  as well as by grant IT1453-22 “Grupos
  Consolidados UPV/EHU  del Gobierno  Vasco”.
\end{acknowledgments}

%\bibliographystyle{prstyown}
%\bibliography{dft,dft_nahual,ownpub,nanodmft}

\end{document}